\documentclass[aps,preprint,onecolumn]{revtex4}
\usepackage{ifpdf}
\ifpdf
\fi

\usepackage[utf8]{inputenc}

\usepackage{graphicx}
\usepackage{epstopdf}
\usepackage{dcolumn}
\usepackage{bm}

\usepackage{SIunits}
\usepackage{subfigure}
\usepackage{hyperref}
\usepackage{amsmath} 
\usepackage{booktabs}

\begin{document}
\title{First Evaluation of Dynamic Aperture at Injection for FCC-hh\footnote{This Research and 
    Innovation Action project submitted to call H2020-INFRADEV-1-2014-1 receives funding from the 
    European Union's H2020 Framework Program under grant agreement no. 654305.}}
\author{B.~Dalena, D.~Boutin, A.~Chanc\'e}
\affiliation{CEA,IRFU,SACM, Centre de Saclay, F-91191 Gif-sur-Yvette, France}
\author{B.~Holzer, D.~Schulte}
\affiliation{CERN, Geneva, Switzerland}

\begin{abstract}
   In the Hadron machine option, proposed in the context of the Future Circular Colliders (FCC) study, 
   the dipole field quality is expected to play an important role, as in the LHC. A preliminary 
   evaluation of the field quality of dipoles, based on the Nb$_{3}$Sn technology, has been provided by 
   the magnet group. The effect of these field imperfections on the dynamic aperture, using the present 
   lattice design, is presented and first tolerances on the b$_3$ and b$_5$ multipole components are evaluated.
\end{abstract}

\maketitle

\section{Introduction}
The main dipole magnets are critical elements for the machine performance in the case of LHC and of FCC-hh.
In particular their field quality impacts the long term stability of the particles in the machine.
The behavior of the particles in presence of magnet imperfections cannot be cured by dedicated 
feedback, therefore, it is important to know them in advance and to correct them if they reduce, below the 
safety limit, the Dynamic Aperture (DA) of the machine, defined as the region in phase space where stable 
motion occurs.
As in the LHC also in the FCC-hh different magnets are expected to play
a role in the definition of the DA at injection and at collision energy.
At injection the main dipole field quality is the major contributor to the reduction of DA 
in LHC~\cite{injLHC}, while at collision the triplet field quality and the beam-beam effects are 
the major sources of the DA reduction~\cite{colLHC}.
The baseline injection energy for FCC-hh has been fixed at 3.3 TeV, which gives more or less the
same ratio between injection and collision energy as in the LHC.
In the following we discuss the first estimate of main dipole field quality and magnet specifications,
using DA as figure of merit.
The optics used for this study are briefly introduced in section~\ref{opt}. In section~\ref{mag} the first 
estimate of dipole field quality together with the arc magnets specifications are presented. 
The first evaluation of tolerances on b$_3$ and b$_5$ using dynamic aperture and detuning with amplitude and 
momentum are reported in section~\ref{da} and~\ref{dt}. Finally, possible improvements of these tolerances 
estimations are qualitatively analyzed in section~\ref{disc}.

\section{\label{opt}Optics}

\begin{figure}[!htb]
   \centering
   \includegraphics*[width=7.cm]{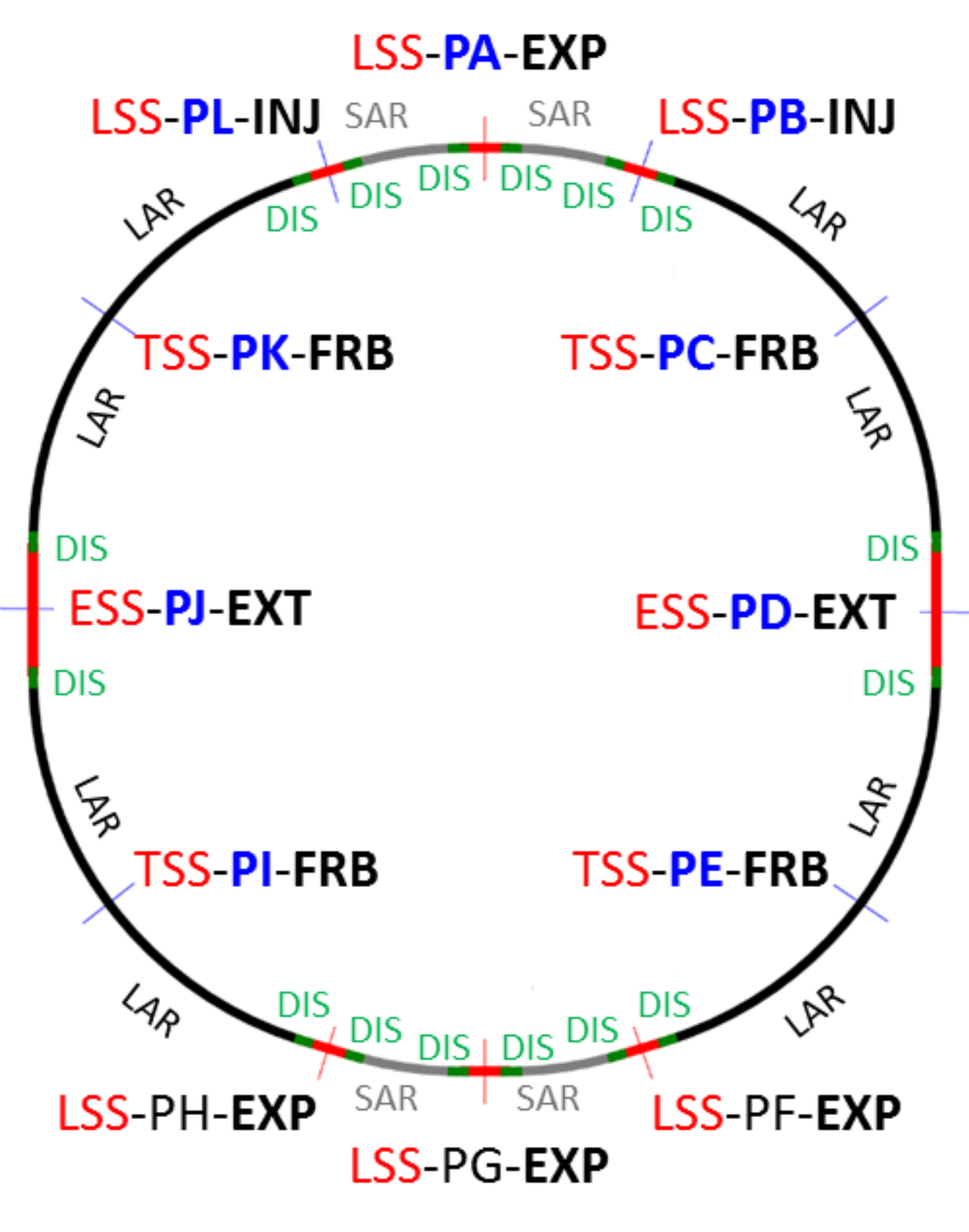}
   \caption{\small Present layout of FCC-hh ring.}
   \label{fig:1}
\end{figure}

The current layout of the FCC-hh ring (see Fig.~\ref{fig:1}) is made of 4 short arcs (SAR), 
4 long arcs (LAR), 6 long straight sections (LSS) and 2 extended straight sections (ESS).
The main ring parameters and the main functionality of each of the insertions are reported 
in~\cite{AntIpac16}.
We have considered different versions of insertion optics for the Interaction Region (IR) and the momentum
collimation, as far as they became available, while keeping the same arc, injection, extraction and betatron
collimation optics. 

Three versions of the IR insertion optics have been considered.
The first version has been designed with a L$^*$ of 36 m~\cite{RomIpac15},  a $\beta^{*}$ of 3.5 m is used for the study of DA
at injection and the ultimate $\beta^{*}$ of 0.3 m at collision (in the following called v2).  
In the second version of the IR optics the $\beta^{*}$ value at injection has been increased to 4.6 m 
to ensure that the aperture bottleneck, and as a consequence the collimation settings, are dominated by the arcs 
and not the IRs (in the following called v4).
Finally, the third version of the IR has been designed with a L$^*$ of 45 m, due to experimental detector
constraints (in the following called v5~\cite{AndyFCC}).
\begin{figure}[!htb]
   \centering
  \subfigure[]{
    \includegraphics[width=0.45\textwidth]{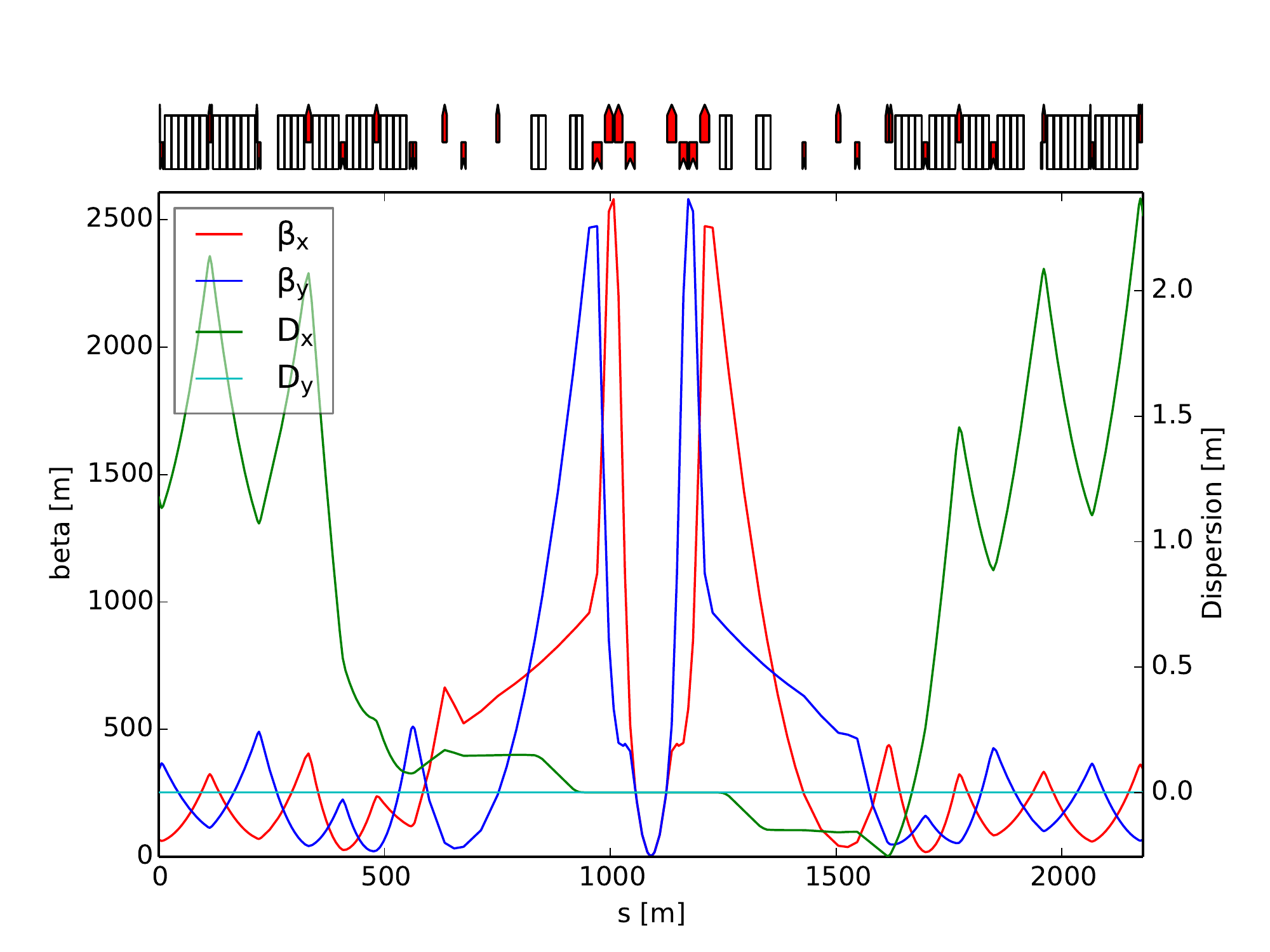}
  }
  \subfigure[]{
    \includegraphics[width=0.45\textwidth]{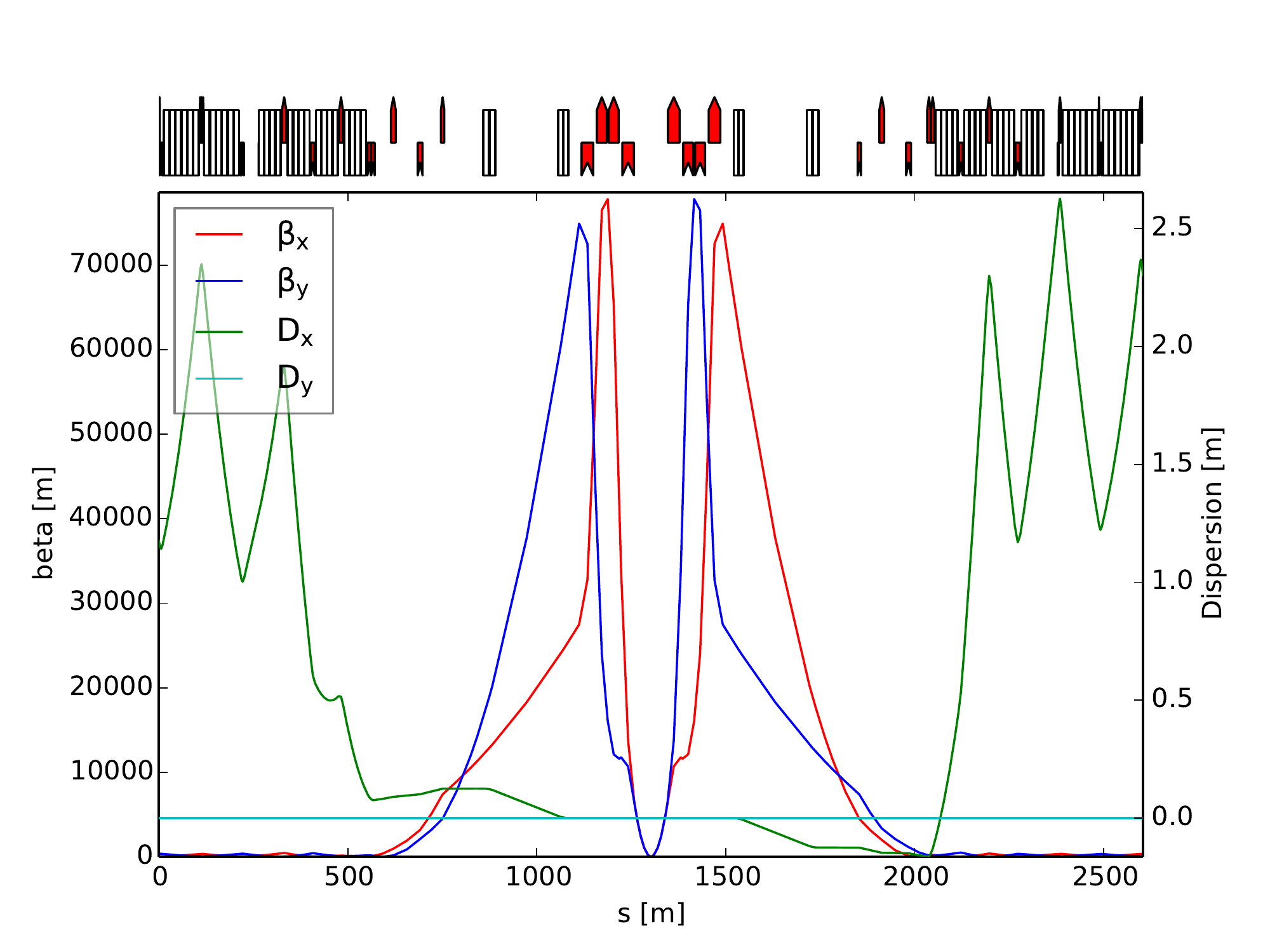}
  }
   \caption{\small Interaction Region optics at injection v4, with L$^*$ 36 m and $\beta^{*}$ 4.6 m (a).
   Interaction Region optics at collision v5, with L$^*$ 45 m and $\beta^{*}$ 0.3 m (b).}
   \label{fig:2}
\end{figure}
Figure~\ref{fig:2} shows the optics function of IR v4 at injection and of IR v5 at collision. 
In the IR v4 and v5 the phase advance between the IPs A and G and the first focusing/defocusing sextupole of 
the SAR is respectively adjusted to 90$^{\circ}$ modulo 180$^\circ$ in the horizontal/vertical plane. 
In all optics versions, the first order chromaticity is corrected ($Q^{'} = 2$) by two sextupole families 
distributed in the SAR and LAR.

Three versions have been used for the momentum collimation optics, as well: the first two are slightly different versions
of simple FODO cells, which have a maximum dispersion of 5 and 4 meters (called fodo90 v1 and fodo90 v2, respectively).
The third version of the optics has been designed by scaling the betatron functions and the lengths from the LHC, 
using the factor $k = \sqrt{\frac{50}{7}}$ (ratio between the nominal beam energy of FCC and of LHC). 
A number of FODO cells with 90$^{\circ}$ phase advance has been added downstream
to fill the 4.2 km foreseen for the ESS insertion region (in the following called lhc v1).  
The concept of the injection and extraction insertions can be found in~\cite{BartIpac15},~\cite{BartFCC16},
and the present betatron collimation design is described in~\cite{FiasIpac16}.

\begin{figure}[!htb]
  \centering
  \subfigure[]{
    \includegraphics*[width=7.cm]{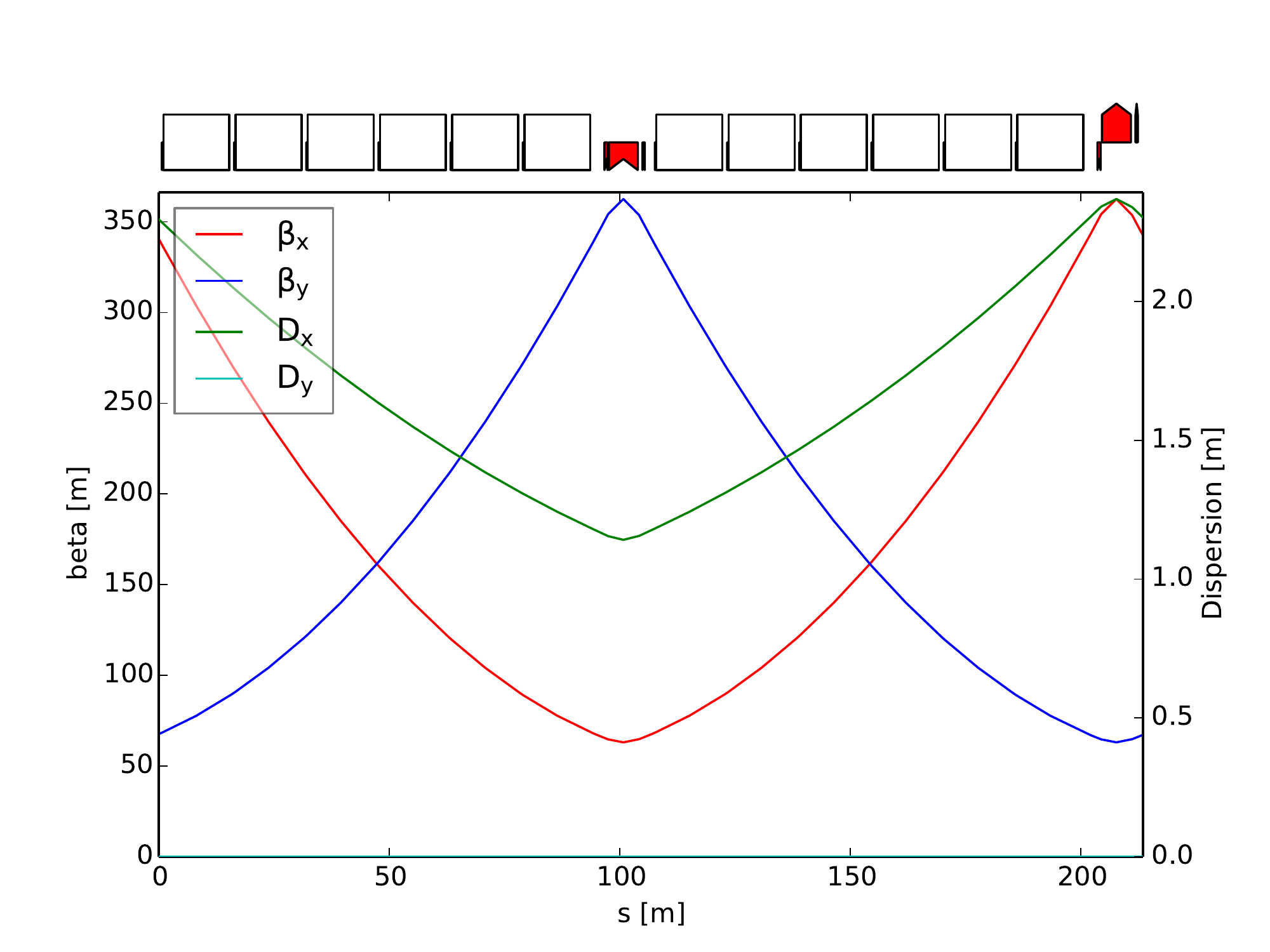}
  }
  \subfigure[]{
    \includegraphics*[width=7.cm]{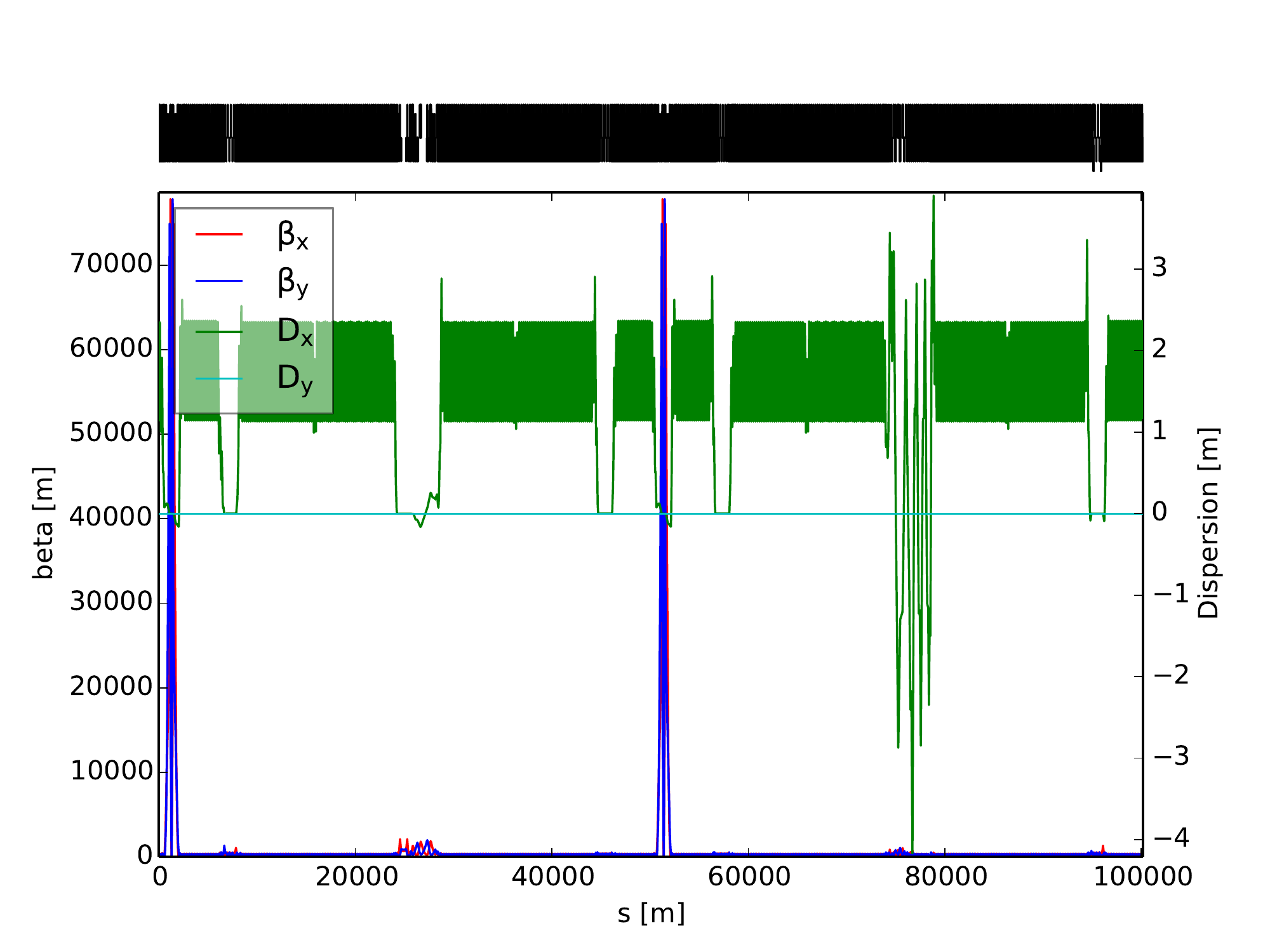}    
  }
  \caption{\small Optics functions of the Arc FODO cell (a) and the whole ring for FCC-hh (b).}
  \label{fig:3}
\end{figure}

The FODO cells of the arcs are optimized to have the largest filling ratio~\cite{Ant15},~\cite{IoIpac15}
(Fig.~\ref{fig:3} (a)). The phase advance in the FODO cells is exactly 90$^{\circ}$ in the SAR whereas it is 90$^{\circ}$ 
$\pm\epsilon_{x,y}$ in the LAR. The value of $\epsilon_{x,y}$ is adjusted to tune the whole ring.
In particular for the optics considered in this study the value of $\epsilon_{x,y}$ changes up to 1$^{\circ}$,
according to the IR and momentum collimation optics version.
A dipole is removed at the middle of the LAR to save some space for the technical straight sections (TSS). 
The optics of the whole ring is shown in Fig.~\ref{fig:3} (b).

\section{\label{mag}Dipole Field Quality and Magnets specifications}

\begin{figure}[!htb]
   \centering
   \includegraphics*[width=14.cm]{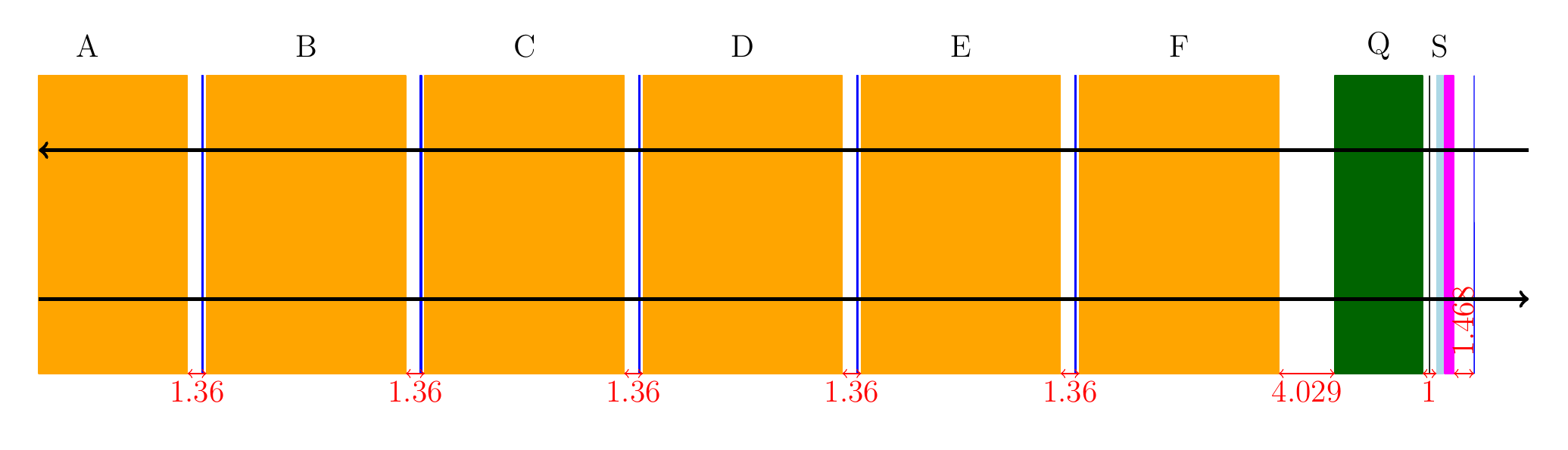}
   \caption{\small Arc half-cell layout.}
   \label{fig:4}
\end{figure}

Currently the total arc cell length is \mbox{$\sim$214~m} with 12 dipoles and 12 spool pieces 
for the correction of the b$_3$ component of the main dipole, one attached to each dipole of the cell 
(as shown in Fig.\ref{fig:4}). 
Both the dipole and the spool piece correctors have the same length of the LHC corresponding magnets. 
The same interconnection lengths between two dipoles, and between the main dipole and the main quadrupoles 
have been estimated to be feasible by the magnet group~\cite{Eziofcc}. 
Octupoles for Landau damping and octupole or decapole correctors are not 
included in the current version of the lattice layout, neither are skew sextupoles. 
The dipoles are set at almost their maximum strength (\mbox{16~T}). At this field and at a reference radius
of \mbox{17~mm} a first estimate of the expected field quality for the injection and the collision energy 
has been provided by the magnets group~\cite{Eziofcc}, as shown in Table~\ref{tab:1}. 
\begin{table}[hbt]
   \centering
   \caption{\small Multipoles used for the main dipole field quality. The values are in units of 10$^{-4}$ at 
     $R_{ref}$=17~mm.}
   \setlength{\tabcolsep}{1pt}
   \begin{tabular}{lcccc}
       \toprule
                       & \textbf{systematic}  &  & \textbf{uncertainty} & \textbf{random} \\
       \textbf{Normal} & \textbf{inj $b_{n_S}$}   & \textbf{col $b_{n_S}$} & \textbf{$b_{n_U}$} & \textbf{$b_{n_R}$}
       \\
       \midrule
           3      & -5    & 20   & 0.781  & 0.781 \\ 
           4      & 0     & 0    & 0.065  & 0.065 \\ 
           5      & -1    & -1.5 & 0.074  & 0.074 \\ 
           6      & 0     & 0    & 0.009  & 0.009 \\
           7      & -0.5  & 1.3  & 0.016  & 0.016 \\ 
           8      & 0     & 0    & 0.001  & 0.001 \\ 
           9      & -0.1  & 0.05 & 0.002  & 0.002 \\ 
       \midrule
       \textbf{Skew} & $a_{n_S}$ & $a_{n_S}$ & $a_{n_U}$ & $a_{n_R}$\\
       \midrule
           3      & 0     & 0   & 0.256  & 0.256  \\ 
           4      & 0     & 0   & 0.252  & 0.252  \\ 
           5      & 0     & 0   & 0.05   & 0.05   \\ 
           6      & 0     & 0   & 0.04   & 0.04   \\
           7      & 0     & 0   & 0.007  & 0.007  \\ 
           8      & 0     & 0   & 0.007  & 0.007  \\ 
           9      & 0     & 0   & 0.002  & 0.002  \\ 
           10     & 0     & 0   & 0.001  & 0.001  \\ 
       \bottomrule
   \end{tabular}
   \label{tab:1}
\end{table}
 
As in the LHC case~\cite{LHCreport}, the magnetic field expansion used for the magnets reads~\cite{Wolf}: 
\begin{equation}\label{eq:1}
    B_y + i B_x = B_{ref} \displaystyle\sum_{n=1}^{\infty} (b_n + i a_n) \left ( \frac{x + i y}{R_{ref}} \right )^{n-1} 
\end{equation}

where $B_{ref}$ represent the magnetic field at the reference radius $R_{ref}$, for the principal harmonics. 
The subscript $n=1$ refers to dipole, $n=2$ to a quadrupole and so on. Each multipole harmonics entering in the dipole 
field expansion ($a_n$ and $b_n$) is modeled as the sum of three contributions:
\begin{equation}\label{eq:2}
    b_n=b_{n_S} + \frac{\xi_U}{1.5}b_{n_U} +\xi_Rb_{n_R} 
\end{equation}

where $\xi_U$ and $\xi_R$ denote the random numbers with Gaussian distribution truncated 
at 1.5 and 3~$\sigma$, respectively. 

The $b_{3}$ harmonics of the dipole field is corrected by the spool pieces.
In order to have the same specification for the $b_{3_S}$ component ($\leq$ 3 units)
as calculated for LHC in Ref.~\cite{Steph}, the maximum strength of the spool pieces correctors required  
is two times the LHC strength value. With present technology, is possible to obtain a gradient of 
\mbox{4430~T/m$^2$} (which is 3 times the LHC strength) 
for a magnet with the same length as the LHC one \mbox{(0.11~m)}~\cite{Eziofcc}.
This ensures the possibility to correct up to 6 units of the systematic b$_3$ component in the main dipoles
at collision energy.\\
The maximum strength used for the main quadrupole of the arcs is equivalent to a field gradient of 360 T/m 
over a 6.3 meter-long magnet. 
A first study of the main quadrupole design shows that a \mbox{400~T/m} gradient is possible over a \mbox{6~m} 
long magnet~\cite{Eziofcc,vedrine}.\\
For the orbit correctors, current technology is able
to provide up to \mbox{4~T} for a 1~meter-long magnet, which seems to be enough given the first evaluation
of orbit correction and mis-alignment tolerances (see ref.~\cite{David} for details).
The only concern, so far, is the required strength of the sextupoles for the ultimate $\beta^*$ of 0.3~m, in fact
with the present technology a gradient of 5560~T/m$^2$ over a magnet length of 0.37 m is feasible, therefore a gradient of 
$\sim$16000 T/m$^2$ can be obtained over a magnetic length of 1.2 m~\cite{Eziofcc}. 
This leaves no margin for reducing $\beta^*$, unless special chromaticity correction schemes are considered.\\
The main specifications of all the magnets, as agreed with magnets group, are summarized in Table~\ref{tab:2}.
\begin{table}[hbt]
   \centering
   \caption{\small Arc magnets specifications.}
   \begin{tabular}{lccc}
       \toprule
                       & \textbf{magnetic}   & \textbf{max}      &                \\
       \textbf{magnet} & \textbf{length [m]} & \textbf{strength} & \textbf{unit}  \\
       \midrule
       dipole          & 14.3    & 16     & T          \\ 
       spool pieces    & 0.11    & 4430   & T/m$^2$ \\ 
       quadrupole      & 6.0     & 400    & T/m        \\ 
       sextupole       & 1.2     & 16059. & T/m$^2$    \\
       orbit corrector & 1.0     & 4      & Tm         \\ 
       dipole-dipole spacing & 1.36 & & m \\
       quadrupole-dipole spacing & $>$3.67 & & m \\
       \bottomrule
   \end{tabular}
   \label{tab:2}
\end{table}

\section{\label{da} Dynamic Aperture}
In this section we analyze the impact of the main dipole field errors, reported in Table~\ref{tab:1},
on the long term stability of the machine (DA). 
The DA has been computed simulating the particles motion over 10$^5$ turns, using a set
of initial conditions distributed on a polar grid, in such a way to have 30 particles (different
initial conditions) for each interval of 2$\sigma$ in a range from 2 to 40. 
Five different phase space angles have been used: 15$^{\circ}$, 30$^{\circ}$, 45$^{\circ}$, 60$^{\circ}$ and 75$^{\circ}$.
Moreover, in all tracking simulations the fractional parts of the tunes 
have been fixed to .28 and .31 at injection and to .31 and .32 at collision, as for LHC upgrade in luminosity. 
Misalignment errors are not included in order to evaluate the effect of the multipole errors alone, 
as also been done in~\cite{injLHC}. As far as the dipole field imperfections are concerned, 
sixty different machines (also called seeds) have been generated, using Eq.~(\ref{eq:2}). 
The $\xi_U$ random number is kept constant for all the dipoles of the same arc, while $\xi_R$
changes for each dipole. The momentum offset is set to \mbox{$7.5\times10^{-4}$} and 
\mbox{$2.7\times10^{-4}$} at the injection and at collision energy, 
same as the LHC, respectively. The normalized {\it r.m.s.} beam emittance 
is kept to $\varepsilon_n=$2.2$\mu$m for both the injection and collision energy (3.3 TeV and 50 TeV). 
In the following, we discuss the results of DA computation at collision first and at injection after.

\subsection{{\it Collision}}
\begin{figure}[!htb]
   \centering
   \includegraphics*[width=8.8cm]{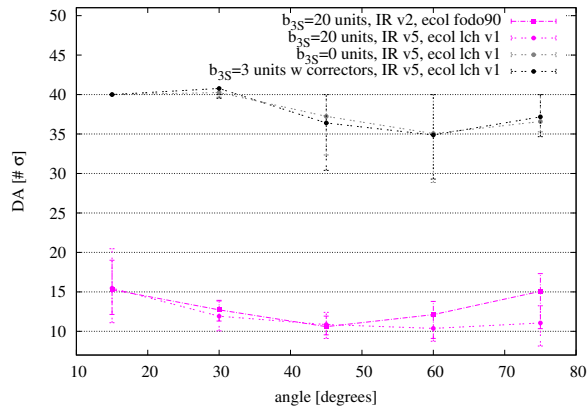}
   \caption{\small Dynamic Aperture (DA) at the collision energy of 50~TeV in number of {\it r.m.s.} beam sizes
     as a function of the initial phase space angles, see text for details.}
   \label{fig:5}
\end{figure}

At collision energy, for the ultimate $\beta^*$ value of 0.3~m the minimum DA without any magnet imperfections 
is  above 54$\sigma$, with minimum values at $30-45^{\circ}$.
When the dipole field imperfections of Table~\ref{tab:1} are included in the simulations, the DA drastically 
drops with a minimum DA value of less than 10$\sigma$, as shown in Fig.~\ref{fig:5}. This is mainly due to the 
geometric aberrations induced by the 20 units of $b_{3_{S}}$ as shown by the grey dots in Fig.~\ref{fig:5}, 
where the $b_{3_{S}}$ value is set to 0.
Moreover, the main sextupole strengths required to correct the chromatic aberrations induced by this $b_{3_{S}}$
are a factor 4 out of the reach of present technology. 
In order to ensure that the arcs have a small impact on the DA at collision (which is already greatly reduced 
by triplet imperfections~\cite{Romanfcc} and beam-beam) it is important to fully correct the $b_{3_{S}}$.   
Furthermore, given the maximum integrated strength reachable by the spool piece correctors, the maximum amount 
of $b_3$ we can correct is $\sim$6 units (as discussed in previous section). The black dots in Fig.~\ref{fig:5} 
show the perfect compensation of the chromatic and geometric aberrations due to 3 units of $b_{3_{S}}$ at collision. 
The average $b_{3}$ of each of the 8 arcs is corrected by the 12 spool pieces attached to each of the main dipoles.
About 54$\%$ of the maximum strength of the spool pieces is used to correct the 3 units of $b_{3_{S}}$.   
A maximum value of 3 units is assigned as target value for $b_{3_{S}}$, which (according to the magnet group) seems 
to be feasible if up to 7 units of $b_{3_{S}}$ are allowed at injection~\cite{Eziofcc}.
The different optics versions considered does not change significantly the DA at collision, as shown by comparing
the pink dots and squares in Fig.~\ref{fig:5}.
 
\subsection{{\it Injection}}
\begin{figure}[!htb]
   \centering
   \includegraphics*[width=8.8cm]{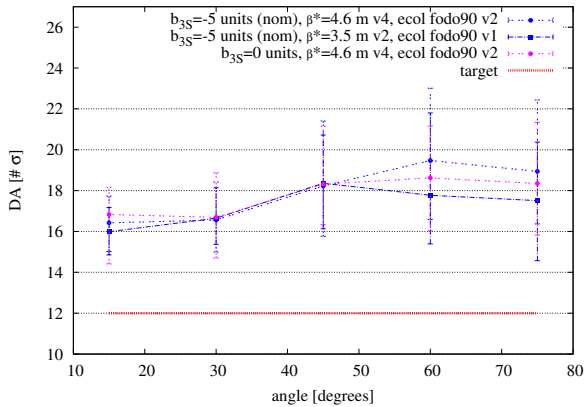}
   \caption{\small Dynamic Aperture (DA) at the injection energy of 3.3~TeV in number of {\it r.m.s.} beam sizes
     as a function of the initial phase space angles, see text for details.}
   \label{fig:6}
\end{figure}

At injection, without dipole imperfections the DA is above 80$\sigma$ for each angle 
explored. As far as the main dipole field imperfections are considered in the tracking simulations 
the minimum DA reduces to 14$\sigma$, as shown by the blue dots and squares in Fig.~\ref{fig:6}. 

In these simulations, the geometric aberrations induced by the sextupole component
of the dipole field are not corrected, while the chromatic aberrations, induced by the  
$b_3$ harmonics of the dipoles, are corrected using the main sextupoles of the arcs.
The minimum DA is above the target value of 12$\sigma$ (safety margin adopted for the LHC design).
Moreover, by comparing the blue dots and squares in Fig.~\ref{fig:6} with the pink dots, where the 
systematic $b_3$ harmonics is set to 0 on purpose, the geometric aberrations generated by the
5 units of $b_{3_{S}}$ are well compensated by the $\sim90^{\circ}$ phase advance of the arc cell.
Moreover, as far as the systematic b$_3$ component of the main dipole field errors stays at 5 units and the 
variation of the arc cell phase advance is around 1$^\circ$, the impact on the DA is negligible as shown by 
comparing blue dots with blue squares in Fig.\ref{fig:6}.  
\begin{figure}[!htb]
   \centering
   \includegraphics*[width=8.8cm]{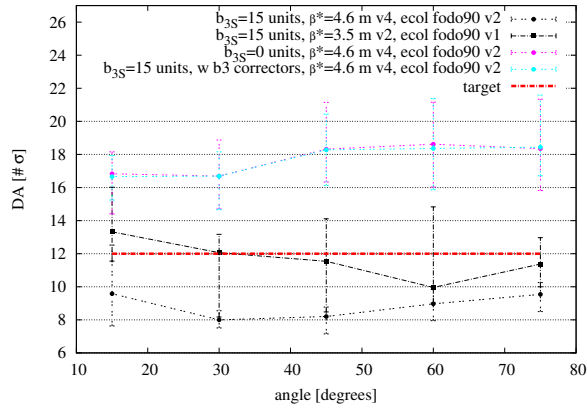}
   \caption{\small Dynamic Aperture (DA) at the injection energy of 3.3~TeV in number of {\it r.m.s.} beam sizes 
     as a function of the initial phase space angles, see text for details.}
   \label{fig:7}
\end{figure}

The black dots and squares in Fig.~\ref{fig:7} represent DA computed with $b_{3_{S}}$ set to 15 units 
on purpose, in order to see visible effects on DA. In both cases the minimum DA is below the target value of 
12$\sigma$ (i.e. $\sim$8$\sigma$). Not only the geometric aberrations 
generated by the $b_{3_{S}}$ are no more perfectly compensated by the arc cell phase advance, but the 
main sextupole integrated strengths, required to correct the chromatic aberrations, run at values well 
above the reach of present technology. Therefore, if the dipole field quality degrades due to persistent 
current at injection energy (up to 15 units of $b_{3_{S}}$), a local correction scheme is needed also at injection, 
namely spool pieces correctors attached to the main dipole, like in the LHC. Moreover, $\sim$1$^\circ$ difference
in the phase advance of the arc cell and the different phases between the arcs start producing significant effects 
on DA ($\sim$ 3$\sigma$ on the average DA at 15$^\circ$), as shown by the black dots and squares in Fig.~\ref{fig:7}. \\  
The light blue dots in Fig.~\ref{fig:7} show that the 15 units $b_{3_{S}}$ are fully corrected using spool 
pieces correctors placed at each dipole of the arcs, correcting each the average $b_{3}$ of the 8 arcs.  
Furthermore, the correctors strength effectively used for the correction is $\sim15\%$ of the maximum
integrated strength that could be reached by present technology~\cite{Eziofcc}, leaving a lot of margin for 
correction.
\begin{figure}[!htb]
   \centering
   \includegraphics*[width=8.8cm]{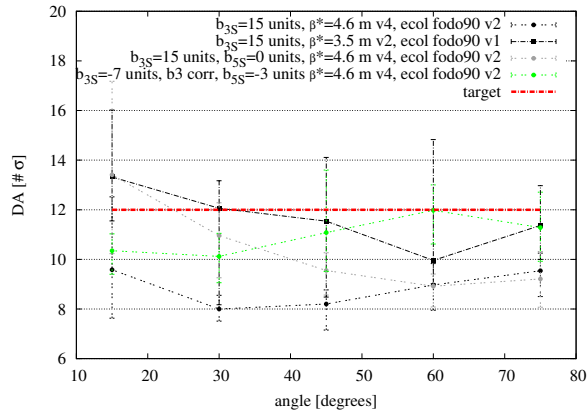}
   \caption{\small Dynamic Aperture (DA) in number of beam $\sigma$ as a function of 
     the phase space angles explored for the baseline injection energy of 3.3~TeV, see text 
     for details.}
   \label{fig:8}
\end{figure}

Finally, the impact on DA of 1 unit of $b_{5_{S}}$ can be seen by comparing the black dots with the grey dots in 
Fig.~\ref{fig:8}, where its value is artificially set to 0. This unit of $b_5$ reduces the average DA of 
$\sim$3$\sigma$ at 15$^{\circ}$, a similar impact is given by the 1$^{\circ}$ difference in the horizontal 
phase advance in the long arc cell
(as shown by comparing the black dots and squares in Fig.~\ref{fig:8}). 
Moreover, with the correction of up to 15 units of $b_{3_{S}}$ the minimum DA is already above the target 
of 12$\sigma$. If the b$_5$ component turns out to be 3 times bigger than the value given in Table~\ref{tab:1},
its impact on DA is not negligible anymore, as shown by the green dots in Fig.~\ref{fig:8} and requires correction. 
In conclusion, if the systematic b$_5$ component of the main dipoles errors stays strictly below 3 units, decapole 
correctors are not required for the present optics design. 

\section{\label{dt}Detuning and Tunes spread}
Multipoles can drive tune shift for particles with energy or amplitude offset. The control of the 
fractional tunes is correlated to the dynamic aperture. In the LHC, the width of the stability island corresponds 
to $\Delta Q = \pm 10^{-2}$, i.e. for an induced tune shift of the order of or bigger than 10$^{-2}$ a visible reduction 
of DA is expected. Detuning with amplitude and with momentum are faster and useful tools to complement DA calculations. 

\begin{figure}[!htb]
  \centering
  \subfigure[]{
    \includegraphics*[width=7.cm]{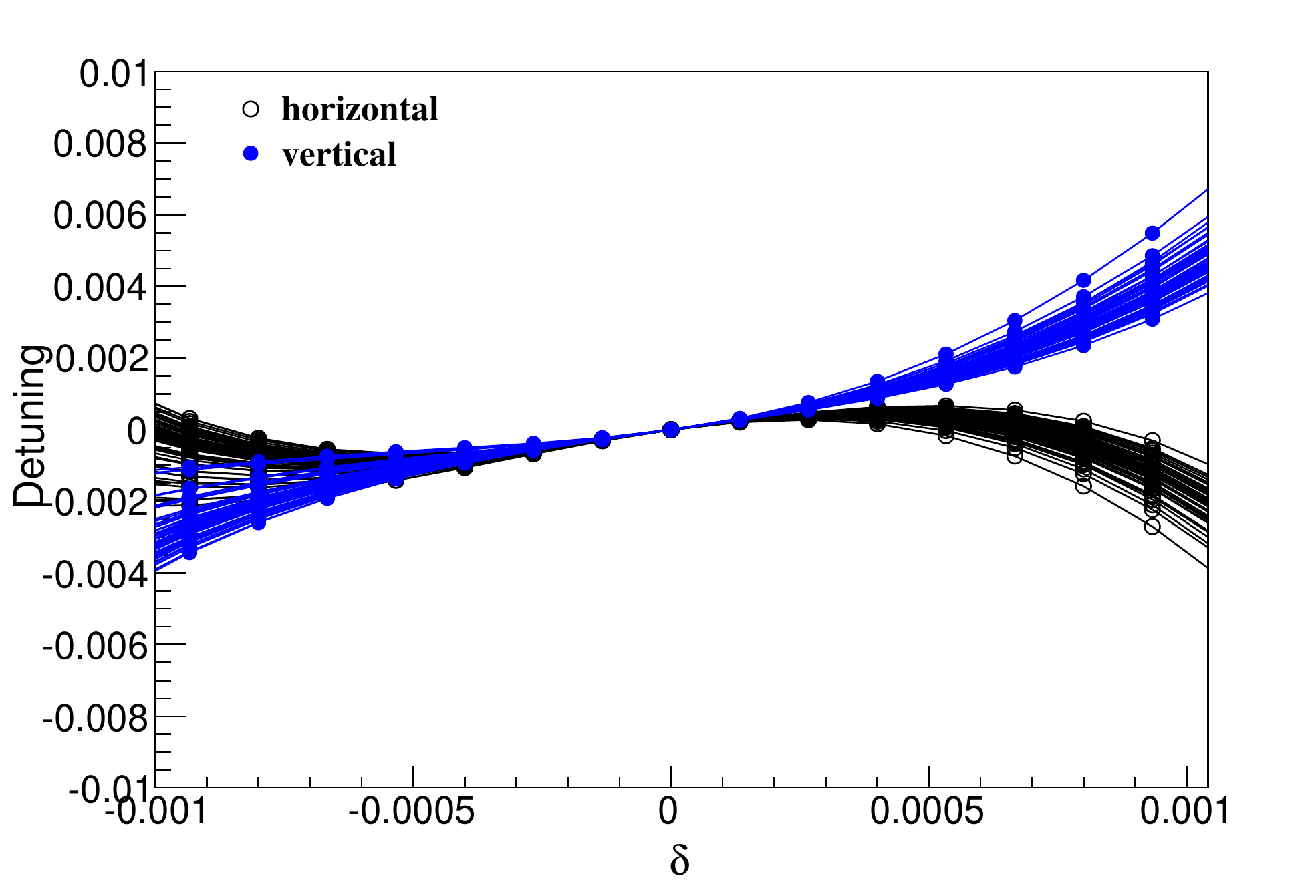}
  }
  \subfigure[]{
    \includegraphics*[width=7.cm]{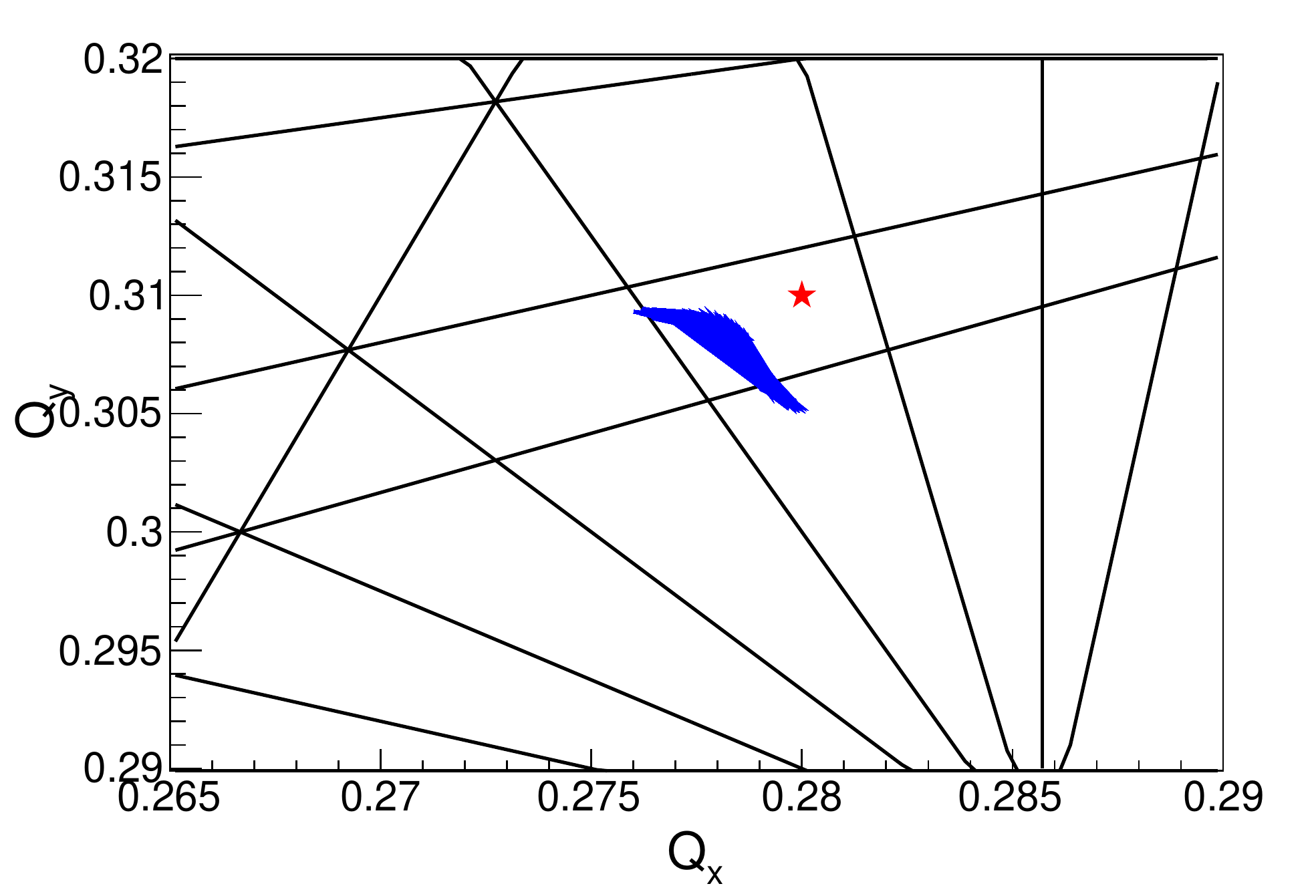}    
  }
  \caption{\small Detuning with momentum (a) and with amplitude up to 7 beam $\sigma$ (b) driven by
    the multipole error Table~\ref{tab:1}, where b$_{3_S}$ has been set to 7 units and corrected with 
    spool pieces (for the 60 seeds used in DA simulations). The lines show resonances up to the 9$^{th}$ order.}
  \label{fig:9}
\end{figure}

In Fig.~\ref{fig:9} detuning with amplitude and momentum are shown in presence of the multipole field errors
reported in Table~\ref{tab:1}, with the only difference that the systematic component of b$_3$ has been set 
to the target value of 7 (current target value) and corrected with the spool pieces.
The chosen working point (the red star in Fig.~\ref{fig:9}) is well away from resonances of order $\leq$5.
The chromatic detuning is 6$\cdot 10^{-3}$ for a maximum $\delta p/p = \pm 10^{-3}$ and the detuning of particles with
amplitude from 1 to 7$\sigma$ (collimator settings proposed for FCC-hh~\cite{FiasIpac16}) is $\sim$5$\cdot 10^{-3}$.
The b$_4$ and b$_5$ multipoles errors are expected to impact the second and third order chromaticity~\cite{SSC}, 
respectively. 
\begin{figure}[!h]
  \centering
  \subfigure[]{
    \includegraphics*[width=7.cm]{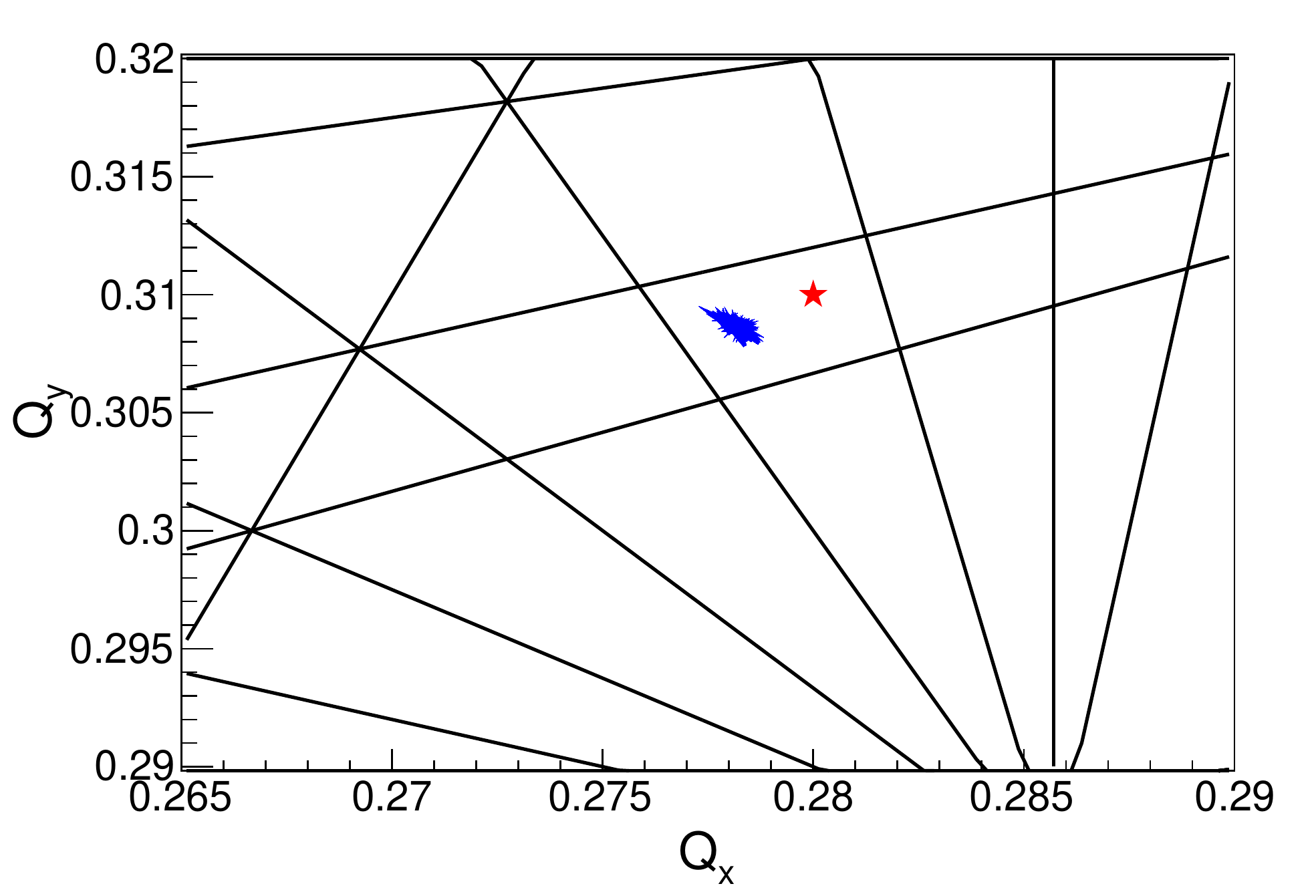}
  }
  \subfigure[]{
    \includegraphics*[width=7.cm]{./figures/footprint_nob2_b3corr_six_2.pdf}    
  }
  \subfigure[]{
    \includegraphics*[width=7.cm]{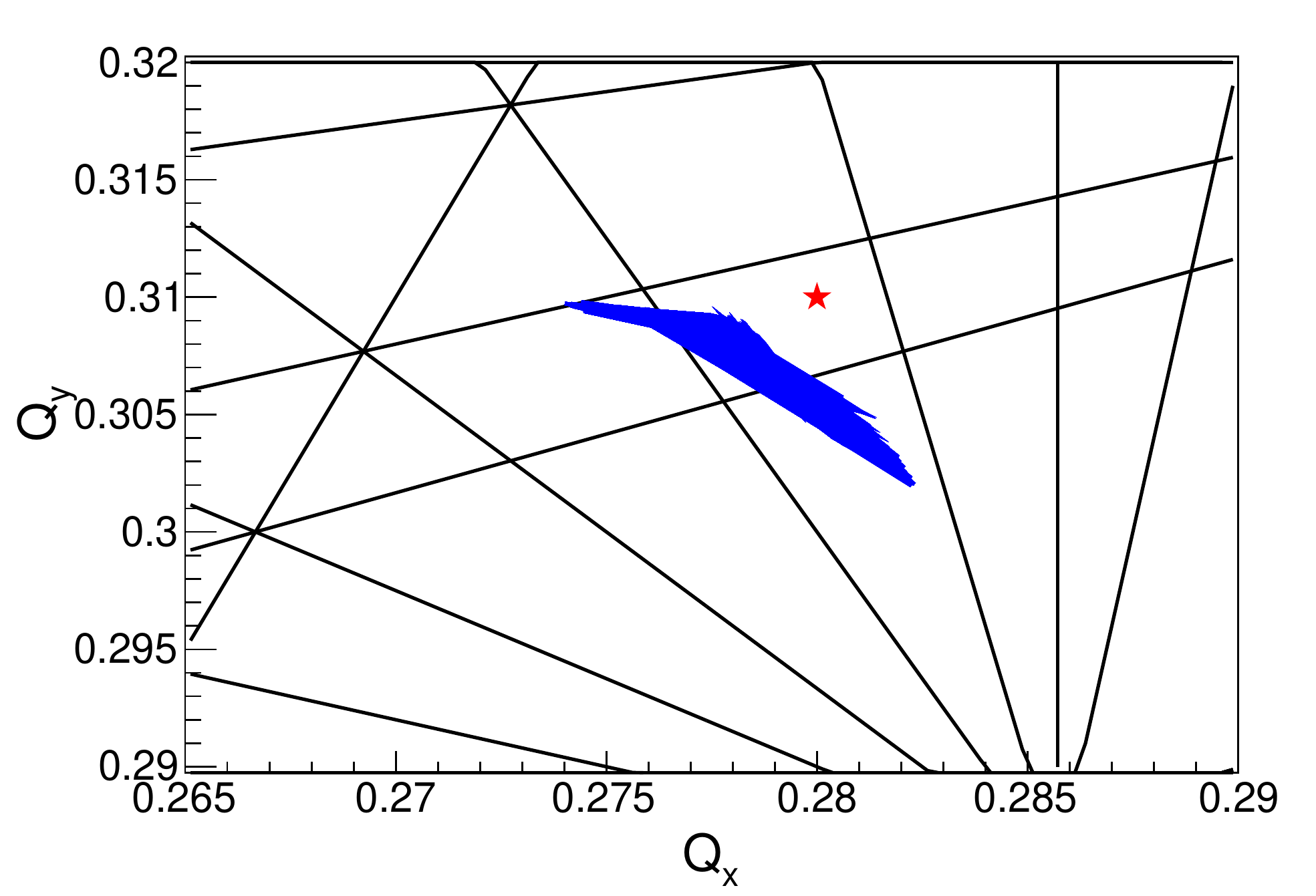}    
  }
  \subfigure[]{
    \includegraphics*[width=7.cm]{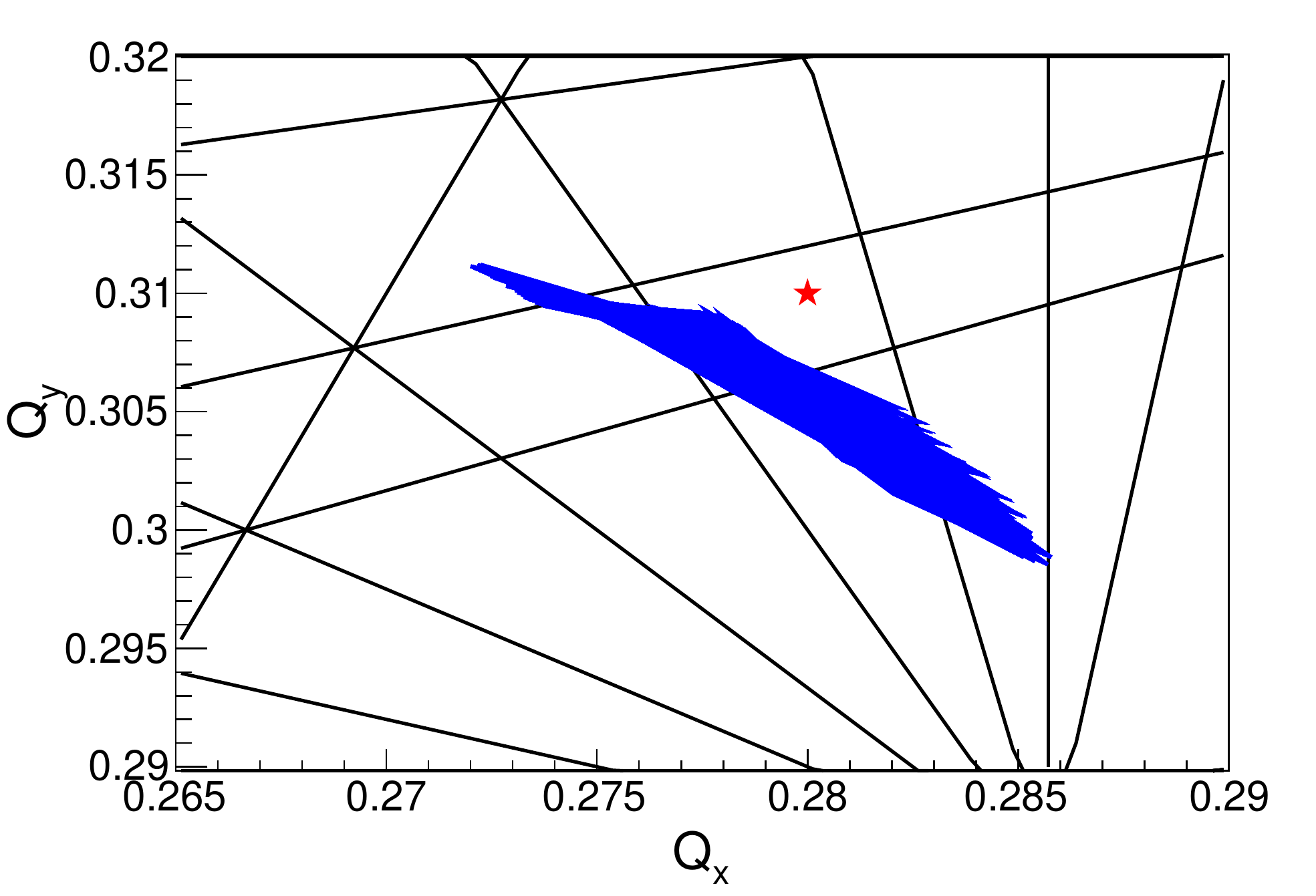}    
  }
  \caption{\small Detuning with amplitude driven by the multipole error Table~\ref{tab:1}, where b$_{5_S}$ has 
    been set to 0 (a), to -1 (b), to -2 (c), and to -3 (d) units, for the 60 seeds used in DA simulations. 
    The b$_{3_S}$ component is set to 7 units and corrected by spool pieces.}
  \label{fig:10}
\end{figure}
\begin{figure}[!h]
  \centering
  \subfigure[]{
    \includegraphics*[width=7.cm]{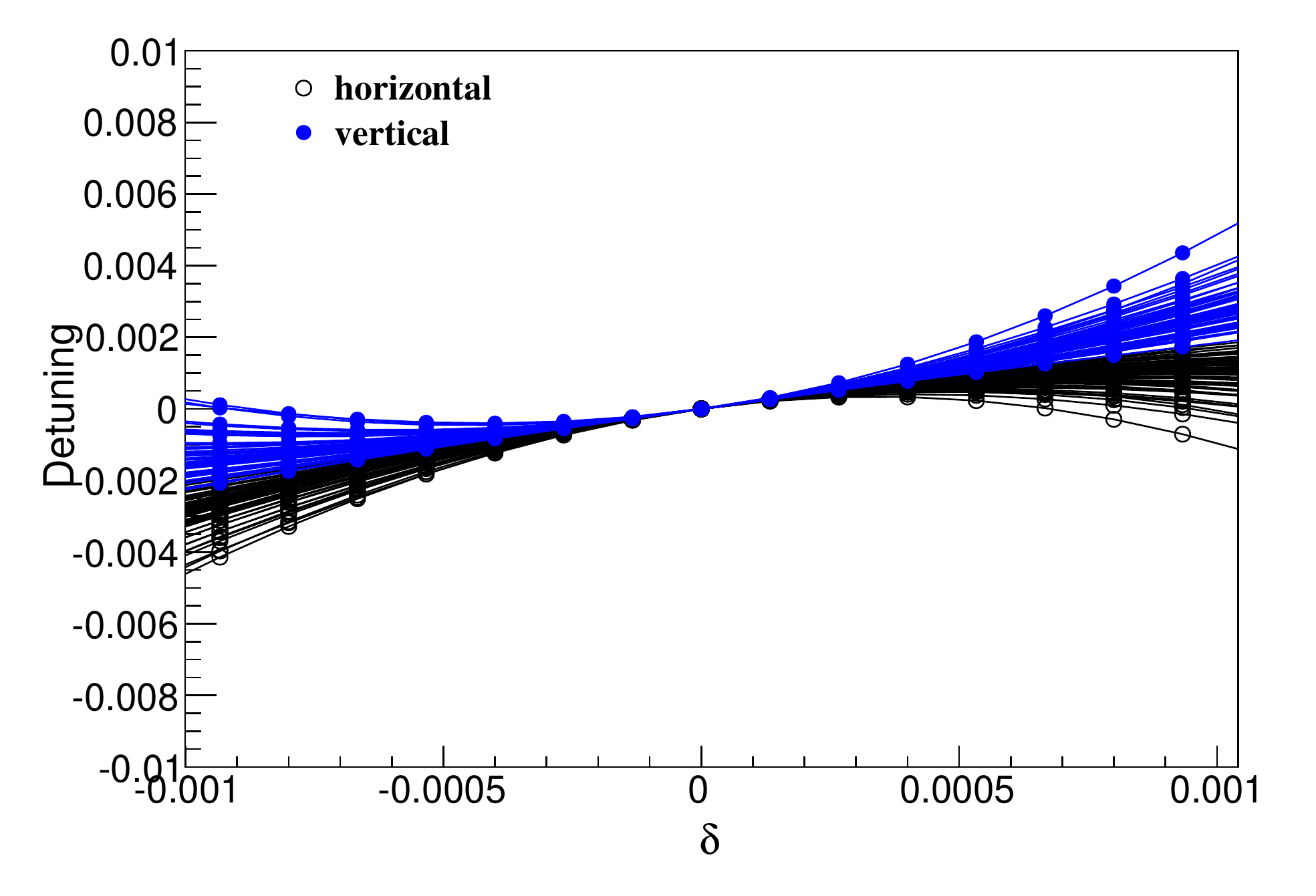}
  }
  \subfigure[]{
    \includegraphics*[width=7.cm]{./figures/tunedp_nob2_b3corr_b5_1_six_nogrid.pdf}    
  }
  \subfigure[]{
    \includegraphics*[width=7.cm]{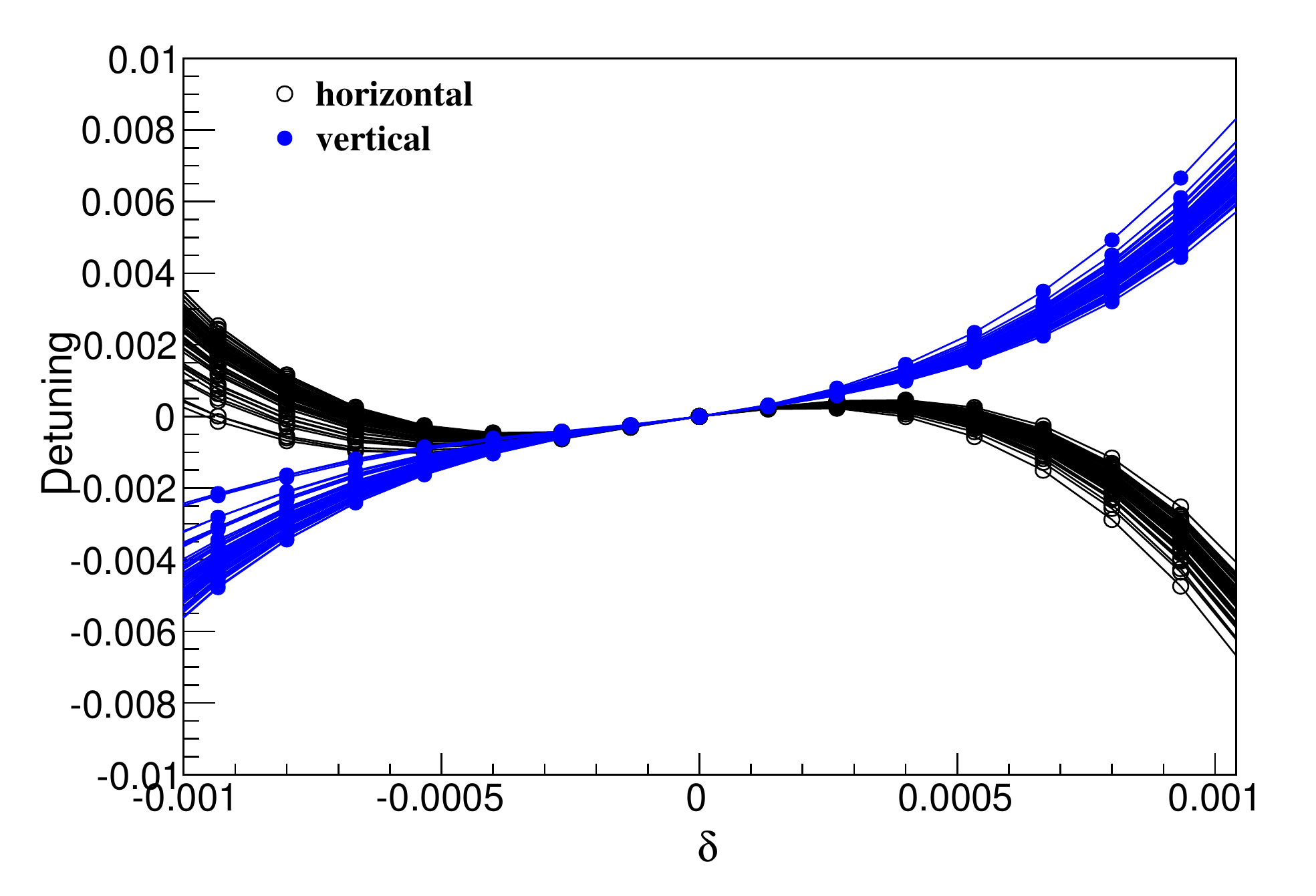}    
  }
  \subfigure[]{
    \includegraphics*[width=7.cm]{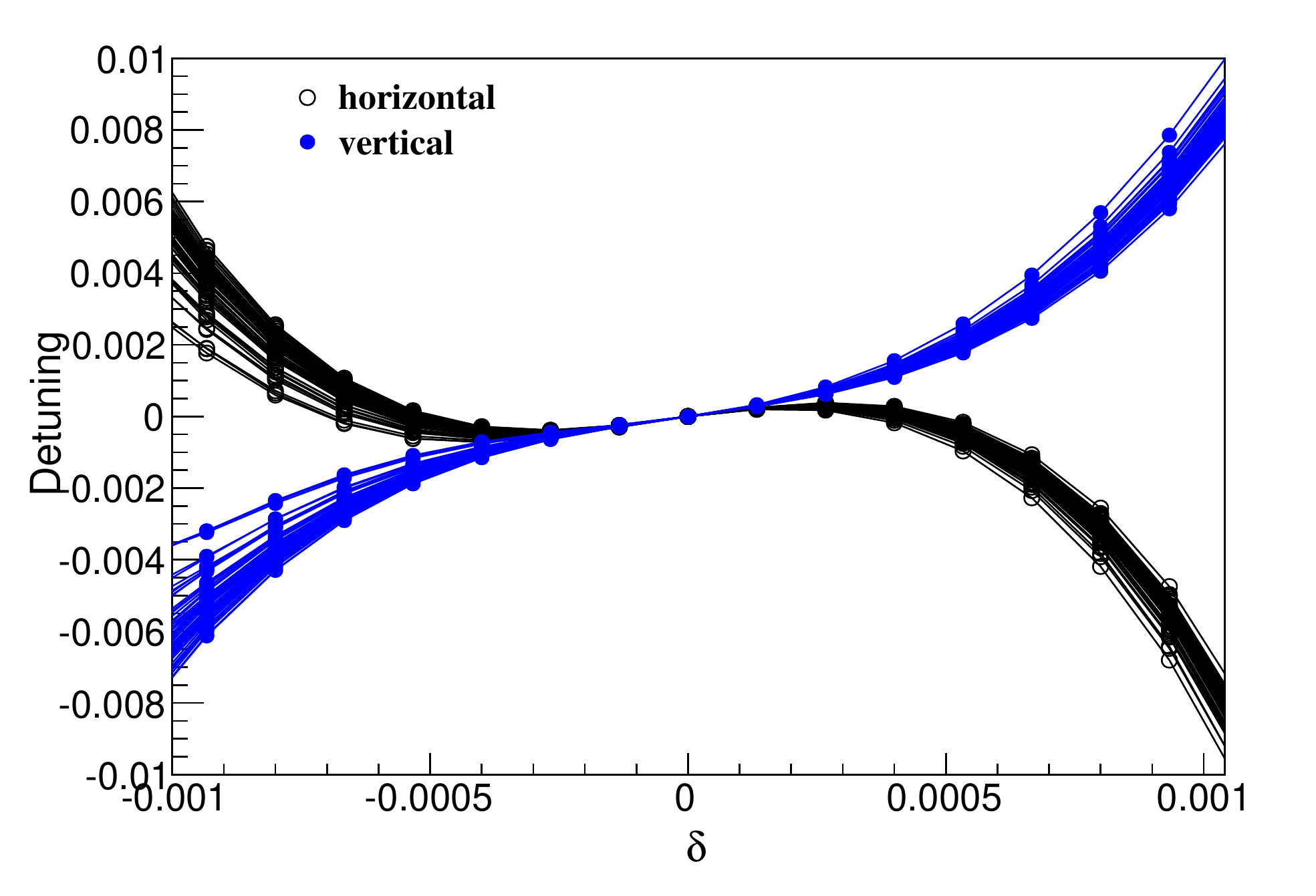}    
  }
  \caption{\small Detuning with momentum driven by the multipole error Table~\ref{tab:1}, where b$_{5_S}$ has been set 
    to 0 (a), to -1 (b), to -2 (c), and to -3 (d) units, for the 60 seeds used in DA simulations. 
    The b$_{3_S}$ component is set to 7 units and corrected by spool pieces.}
  \label{fig:11}
\end{figure}
Figures~\ref{fig:10} and~\ref{fig:11} show the effect of the systematic component of b$_5$ on the detuning with 
amplitude and with momentum, for different values of $b_5$. Comparing these tune shifts with the DA simulations,
shown in Fig.~\ref{fig:8}, we see that a tune shift of 0.9$\cdot$10$^{-3}$ at $\pm 10^{-3}$ gives visible effects 
on DA, as well as a detuning with amplitude of 10$^{-2}$, thus they should be avoided. Therefore, a tolerance on
the maximum b$_5$ component of the field can be fixed to $\leq$ 2 units.
Finally, the other components of the dipole field errors reported in Table~\ref{tab:1} generate all together a tune 
spread well below the value 10$^{-2}$, so they have a minor impact on DA.

\section{\label{disc}Discussion}
In this section further qualitatively considerations are given about the injection energy, about feed-down effect 
and about the impact of dipole type on DA simulations.

\subsection{{\it Injection Energy}}
As far as DA is concerned, the present estimate of the field errors reported in Table~\ref{tab:1} and also the 
new target value for the systematic component of b$_3$ (7 units) at injection guarantee that DA is above the 
target of 12$\sigma$. Considering that the ratio of DA at two different energies is equal to the ratio of the 
$\sqrt{\gamma}$, the lower limit for injection energy, as far as DA is concerned and using the present field 
quality table, is set to $\sim$2.6 TeV. In order to reach the 1.5 TeV proposed in~\cite{injfcc} 
tighter tolerances on the random and uncertainty components of the main dipoles imperfections will be 
required to ensure the DA is above the 12$\sigma$ target value.
Alternatively, a target for DA closer to the collimation settings limit (7 $\sigma$) should be considered.

\subsection{{\it Feed-down effect}}
All tracking simulations showed in this paper do not take into account mis-alignment of the magnetic elements.
This is expected to change the value of the multipoles shown in Table~\ref{tab:1} due to feed-down effects of higher 
order multipoles on the multipoles of lower order.
Following Ref.~\cite{Steph} the expected feed-down of b$_3$ on b$_2$ and of b$_5$ on b$_4$ due to
position errors of main dipoles and spool-piece can be calculated analytically. 
Considering the same alignment position errors reported in 
Table~8 of Ref.~\cite{Steph} one gets for the systematic component b$_{2_S}$:
\begin{equation}\label{eq:3}
    b_{2_S}^{feed-down} = \pm 0.0141 \cdot b_{3_S}  
\end{equation}

which corresponds to values between 0.099 and 0.211, taking b$_{3_S}$ equal to 7 or 15 units, respectively.
For the random component of b$_2$ one gets:
\begin{equation}\label{eq:4}
    \sigma_{b_{2_R}}^{feed-down} = \sqrt{ 0.0043 b_{3_S}^{2} + 0.0058 \sigma_{b_{3_R}}^2 }   
\end{equation}

which gives values between 0.46 and 0.98, taking b$_{3_S}$ equal to 7 or 15 units, respectively.
In the case of 15 units the part of b$_2$ due to feed-down becomes bigger than the estimated error due to the 
main dipole (0.481 units for both the b$_U$ and the b$_R$ component). 
These values should be added to the b$_2$ of the main dipoles and will contribute to the $\beta$-beating,
which impacts the mechanical aperture of the machine. Moreover, main quadrupole, main sextupole and straight
section field errors and mis-alignment will also contribute to the $\beta$-beating and further reduce tolerances on 
acceptable total b$_2$ and a$_2$, and as a consequence of b$_3$, coming from the main dipoles. 

Always following Ref.~\cite{Steph} the systematic and random b$_5$ feed-down on b$_4$ is calculated as: 
\begin{equation}\label{eq:5}
    b_{4_S}^{feed-down} = \pm 4 \cdot 0.0071 \cdot b_{5_S}  
\end{equation}
\begin{equation}\label{eq:6}
    \sigma_{b_{4_R}}^{feed-down} = 4. \cdot \sqrt{ 0.0033 b_{5_S}^{2} + 0.0015 \sigma_{b_{5_R}}^2 }   
\end{equation}

which are equal to 0.028 units and 0.23 for the b$_5$ component reported in Table~\ref{tab:1}.
These values of b$_4$ are bigger than the components reported in Table~\ref{tab:1}, thus they
will have a non negligible impact on DA and can further reduce the tolerance on the b$_5$ component.

\subsection{{\it Dipole type}}
In all the simulations results presented here curved dipole (SBEND type) have been considered, and the multipole field 
expansion is done around the magnetic center of the dipole, which coincide with the reference particle trajectory.
Due to the properties of the Nb$_3$Sn material, it has been proposed to build them straight (RBEND type). 
This implies that the reference particles does not travel always on the magnetic center of the element. The 
maximum beam excursion (the Sagitta $x_S$) due to the dipole field is about 2.5 mm. Therefore, assuming the multipole 
field expansion (of Eq.~\ref{eq:1}) is done always with respect to the magnetic axis. For a systematic offset equal 
to the Sagitta the maximum error on the b$_3$ component of the field can be calculated using the same formula given 
for the feed-down harmonics (see Eq.~(15) of Ref.~\cite{Steph}):

\begin{equation}\label{eq:7}
    b_{3err}=b_{5} \cdot 6 \cdot \left ( \frac{x_s}{R_{ref}} \right )^2 = (b_{5S} + b_{5U} + b_{5R}) 
    \left ( \frac{x_s}{R_{ref}} \right )^2 
    \sim  b_{5S}\cdot0.13 = -0.13 
\end{equation}
    
This value is less than the random component of b$_3$ reported in Table~\ref{tab:1}, which dominate the DA results 
as discussed in section~\ref{da}.
Moreover, considering that the multipole errors in the magnet are dominant at the magnet ends and almost zero at the 
longitudinal center of the magnet, the error on b$_3$ is even less than the previous calculated value, if the beam is 
off-axis in the longitudinal center of the magnets. On the other hand the useful aperture of the beam will be reduced, 
therefore a balance should be found on the beam excursion along $z$ in the magnet.
Finally, in any case the error would be systematic for each of the dipole, thus the error on b$_3$ would be corrected 
by the spool pieces.

\section{Conclusions}
The arc magnet specifications are reported and discussed in connection with the magnet group feedback.
As far as the first estimate of the main dipole field quality is considered, the dynamic aperture of the 
current optics design of the future hadron-hadron collider is above the target value of 12$\sigma$ at injection
energy (3.3~TeV). The current
main dipole imperfections would allow to reduce injection energy to a minimum value of $\sim$2.6~TeV, as far as
dynamic aperture is considered as criterion. The possibility to correct up to 15 units of the systematic $b_3$ 
harmonics of the main dipoles is shown, leaving a lot of margin in the correctors strength at the injection 
energy of 3.3 TeV. A first tolerance on the b$_3$ component is fixed to 7 units at injection but the feed-down 
effect on b$_2$, which impact the $\beta$-beating can further reduce this tolerance.  
No specification for the decapole correctors are given at this stage of the design and a tolerance to  
$\sim$2 units is fixed for the b$_5$ component at injection, but first order feed-down of b$_5$ on b$_4$ can
further reduce this tolerance.
At collision energy the first estimate of the main dipole field quality strongly reduces DA
(below 10$\sigma$). In particular, a new systematic value of the $b_3$ harmonics (3 units) has been specified 
as target.
Moreover, a minimum spool pieces integrated strength of a factor 2 higher with respect to LHC is required.   
Further studies are needed to fully specify the main dipole field quality, in particular evaluate the impact 
on dynamic aperture of feed-down of the main harmonics due to mis-alignment errors. 
\section*{ACKNOWLEDGMENT}
The authors would like to thank Ezio Todesco for the tight collaboration in the definition
of the magnet specifications and St\'ephane Fartoukh for the useful comments and discussions.


\end{document}